\documentclass[groupedaddress,nofootinbib]{revtex4}  
\usepackage{graphics}
\def\x{\vec{x}}
\def\c{\hat{c}}
\def\s{\hat{s}}
\def\partd#1#2{{\frac{\partial #1}{\partial #2}}}
\begin{document}
\voffset=0.25in
\def\page{\vfill\eject\voffset=1in}
\title{Dimension-splitting for simplifying diffusion in lattice-gas models}   

\author{Raissa M. D'Souza}
\email{raissa@bell-labs.com}
\affiliation{Bell Laboratories, Murray Hill, NJ 07974}
\author{Norman H. Margolus}
\email{nhm@mit.edu}
\affiliation{Center for Computational Science, Boston University, Boston,
Massachusetts 02215\\ Artificial Intelligence Laboratory,
Massachusetts Institute of 
Technology, Cambridge, Massachusetts 02139}
\author{Mark A. Smith}
\email{smith2@fas.harvard.edu}
\affiliation{Department of Molecular and Cellular Biology, Harvard
University, Cambridge, Massachusetts 02138\\
New England Complex Systems Institute, Cambridge, Massachusetts 02138} 
\date{\today}

\begin{abstract}
We introduce a simplified technique for incorporating diffusive
phenomena into lattice-gas molecular dynamics models.  In this method,
spatial interactions take place one dimension at a time, with a
separate fractional timestep devoted to each dimension, and with all
dimensions treated identically.  We show that the model resulting from
this technique is equivalent to the macroscopic diffusion equation in
the appropriate limit.

This technique saves computational resources and reduces the
complexity of model design, programming, debugging, simulation and
analysis. For example, a reaction-diffusion simulation can be designed
and tested as a one-dimensional system, and then directly extended to
two or more dimensions. We illustrate the use of this approach in
constructing a microscopically reversible model of diffusion-limited
aggregation as well as in a model of growth of biological films.


\end{abstract}
\maketitle

{\bf Key words:} Dimension-splitting, diffusion, lattice-gas automata,
transport theory, alternating direction method, microscopic
reversibility, parallel computing, fractional timestep method.

\section{Introduction}
In computer modeling of physical systems, there is not always a direct
correspondence between simple models and computationally efficient
ones.  What seems like a direct and intuitive description to a human
does not necessarily correspond to an algorithm that performs well on
a computer.  Simplicity and directness can, however, be valuable
assets regardless of whether they result in better performance.
Simple models are easier to analyze and to translate into correct
computer programs.  In addition, complex models composed of
well-understood simple component models are easy to design and test.
Of course what we would really like is to combine simplicity and
performance.

Molecular dynamics---numerically simulating the classical or
semiclassical behavior of a collection of atoms or molecules---is an
example of a direct technique with wide applicability that is
conceptually simple.  The technique is often used in situations where
a computational solution based on differential equations is infeasible
due to the complexity of the physical dynamics or the complexity of
the boundary conditions.  Often, in molecular dynamics simulations,
the microscopic dynamics of the particles is simplified in order to
enable more computationally efficient and hence larger-scale
simulations, as in the dissipative particle dynamics and lattice-gas
approaches\cite{Cov-diss-prtcl,BU-lga-meeting}.  In a lattice-gas
automaton (LGA), for example, only a small number of particle
velocities are allowed, and particles are confined to the sites of a
finite spatial lattice.  Steps in which particles concurrently hop
between lattice sites alternate with steps in which the particles
present at each lattice site interact.  This simplification of the
microscopic details results in models that combine simplicity and
performance,
while still capturing the full essence of phenomena of interest.

LGA-like models have long been studied in physics as theoretical
models which capture some essence of a physical phenomenon.  The
classic example is the random walk on a lattice, which has been used
as a microscopic model of diffusion\cite{Weiss-rwalks}.  The first
true LGA models were introduced for the theoretical study of fluid
flow\cite{HPP-pra76}.  It was realized only much later that the
locality, uniformity and spatial regularity of models based on LGA's
make them ideal candidates for large-scale simulation on parallel
hardware\cite{FHP-prl86,MTV-prl86,cam8-waterloo}.  Since then, LGA
simulations have been used to study a variety of physical phenomena,
including fluid dynamics, chemical reactions, and changes in
phase\cite{RothZales-book,Rivet-Boon-lgas-book,Dieter-lgas-book,Kapral-prl91,LGA-reactive-phys-report96}.
More generally, lattice dynamics which emulate the locality and
uniformity of physical dynamics occupy a borderland between abstract
theoretical models and practical algorithms: they are often
both\cite{cambook,Smith-phd,Dsouza-phd,Margolus-feynlec}.  The
computer model is the same as the theoretical model, unlike models
based on partial differential equations (PDE's), for example, where an
intervening stage of numerical analysis needed.  Thus lattice models
and related modeling techniques have both a theoretical and a
practical aspect, providing simple models that we can directly
experiment with, visualize and analyze mathematically.

In this paper, we introduce a simple technique for incorporating
diffusive phenomena into LGA and other lattice models.  This technique
is closely related to the classical random walk model, but differs in
important respects.  First, it incorporates exclusion: only a limited
number of particles can occupy each lattice site.  The technique we
use in this manuscript to implement diffusion with exclusion was first
introduced by Toffoli\cite{cambook}.  A second difference involves
randomness: it is explicitly provided by concurrently simulating a
different invertible LGA as a source of stochasticity.  This approach
has also been used before in other parallel LGA
models\cite{cam8-waterloo,Smith-phd}.  Finally, perhaps the most
important distinction from a classical random walk model is that this
technique splits each $n$-dimensional simulation time-step up into $n$
fractional time-steps, updating only one dimension during each
fractional step.  This is similar to some dimension-splitting and
fractional-time-step techniques used in solving
PDE's\cite{Ames,prng-ran2}, but here we apply it in the context of a
lattice-gas with exact particle conservation.
Earlier Cellular Automata (CA) models have used
dimension-splitting\cite{cam6-rotation,Fredkin-note}, but not for
simulating diffusion, and not as a general modeling technique.  For a
review of other techniques for simulating diffusion with discrete
lattice systems see Ref.~\cite{Band-cadiff}.

LGA models of complex materials and phenomena can be constructed in a
modular fashion by combining simpler models---something which is
difficult to do with differential equation based models.  Qualitative
models are particularly easy to construct in this manner, but
realistic quantitative models of physical phenomena can also be
constructed.  For example, to reproduce known chemical behavior,
several subsystems can be combined.  Each subsystem would consist of
independently diffusing chemical species, and interaction between
subsystems ({\em e.g}, chemical reaction steps) would occur in between
independent species-diffusing
steps\cite{Kapral-prl91,LGA-reactive-phys-report96}.  The diffusing
steps can be handled with a dimension-splitting approach, resulting in
a very simple treatment of the diffusion component of the model, which
requires only a small amount of state information at each lattice
site.  A smaller amount of state not only allows larger simulations to
be achieved with given memory resources, but also makes it easier to
couple component subsystems together.  The uniform treatment of
one dimension at a time provides an additional benefit.  Consider, for
example, a chemical reaction model of the sort described above.  Such
a model can be developed and tested as a two-dimensional model and
then converted into a three-dimensional model by simply applying the
same dynamics to an additional dimension.

In general, LGA models have a different domain of applicability than
PDE-based models.  Since they are based on a kind of molecular
dynamics optimized for exact digital computation on massively parallel
hardware, LGA's can be constructed to handle complex interactions and
complex boundaries---circumstances poorly suited to PDE-based
simulations.  For example, running a fluid LGA simulation through an
MRI reconstruction of a real porous rock involves no more
computational work than running it with any other boundary condition.
An example illustrating the power of the LGA approach for complex
systems with complicated boundaries is a lattice-gas model of
amphiphilic fluid flow in porous
media\cite{Cov-etal-porous}.\footnote{Note that in
Sec.~(\ref{subsubsec:LBA}) we introduce the lattice-Boltzmann equation
(LBE) for our model and discuss some of the distinctions between the
LGA and the LBE approaches.}

In this paper we first review a version of Toffoli's LGA model of
diffusion with exclusion.  We follow an analysis due to Chopard and
Droz\cite{ChopDroz-diff,ChopDroz-book} to show explicitly how to start
with the microscopic model and recover the appropriate macroscopic
diffusion equation in the continuum limit.  We then introduce a
dimension-splitting technique for use in LGA models that involve
diffusion.  We show explicitly how to map this dimension-reduced model
onto Toffoli's LGA model.  We then discuss the origins of the
conservation laws present in the model.  Finally we illustrate the use
of this technique as a component for building two composite LGA
models: a microscopically reversible model of diffusion-limited
aggregation, and a model of growth of biological films.  These two
models were developed in two dimensions but have also been run and
studied in three dimensions.

\section{Diffusion}

Modeling diffusion is a cornerstone for building general models of a
wide range of physical phenomena. Examples include models for
reaction-diffusion processes, pattern formation, and heat
flow\cite{cambook,BLS-transportbook,Walgraef-patformbook}.  Models of
diffusive transport alone encompass such diverse phenomena as flow
through porous media\cite{RothZales-book}, flow of fluids in
biological systems\cite{Berg-rwalks}, and flow in geological
processes\cite{RothZales-book,BLS-transportbook}.
 
Two standard approaches to modeling diffusion come from opposite
extremes: the continuum limit, where diffusion is modeled by a partial
differential equation; and the microscopic limit where diffusion is
modeled as a collection of particles undergoing random walks (which is
meant to capture the phenomenon of Brownian motion).  One-dimensional
diffusive phenomena were first described from a macroscopic perspective
by Fick\cite{Crank}.  The generalization to multiple dimensions is the
well known diffusion equation:
\begin{equation}\label{eq:ficks}
\partd{\rho(\x,t)}{t} = D \nabla^2 \rho(\x,t).
\end{equation}
Here $\rho(\x,t)$ is density as a function of space and time, and $D$
(the coefficient of self-diffusion) describes the diffusion rate.

We are interested in modeling diffusive phenomena beginning at the
microscopic scale and recovering Eq.~(\ref{eq:ficks}) in the
appropriate limit.  We first consider a lattice-gas automata (LGA) model of
particles hopping at random along the sites of a periodic lattice. The
particles execute simultaneous random walks while obeying an exclusion
principle.  This model was introduced by Toffoli,
and it was extended and studied extensively by Chopard and
Droz. They show that in the limit of
the infinite lattice, the microscopic particle model maps directly
onto the macroscopic continuum description, Eq.~(\ref{eq:ficks}).
Additionally, they show that the Green-Kubo relation for expressing
transport coefficients (such as the diffusion coefficient $D$) in
terms of autocorrelation functions is recovered exactly.

Below, we present the model and review the basic analysis of Chopard and
Droz. Then we present a new model which introduces a
dimension-splitting technique. Then we show, by a remapping, that both
models are equivalent.

\subsection{An LGA model of diffusion with exclusion}\label{subsubsec:define-model}
Consider a two-dimensional square lattice.\footnote{A similar argument
would apply to models defined on the two-dimensional triangular
lattice, with the transport split into a sequence of three operations
performed one principle lattice direction at a time.}  We can think of
particles hopping between adjacent sites of the lattice with unit
velocity. The particles are initialized on lattice sites so at every
integer time, the particles occupy a lattice site.  The particles can
move in one of four directions: north, west, south, or
east---corresponding to the four lattice directions.  In the language
of lattice-gases, we consider a ``transport channel'', $N_i(\x,t)$,
along each lattice direction, $\c_i$, at each site $\x$ at time $t$,
where $i\in\{0,1,2,3\},$ and $\c_i$ is the unit vector along that
lattice direction.  The transport channel can either be occupied (with
a particle to be moved) or empty: $N_i(\x,t)=1$ or $0$ respectively.
As there are four channels at each site, there can be up to four
particles at each site.

The dynamics can be decomposed into two portions: the ``interaction''
phase and the ``streaming'' phase.  We denote the state {\it after}
interaction but {\it before} streaming as $N'_i(\x,t)$.  Note, the
interval for the completion of both phases is considered the unit
interval of time.  During the streaming phase the particles are
transported along the lattice. If there were no interaction phase, the
particles would stream along the lattice ballistically, with their
velocities never changing.

To implement simultaneous random walks, during the interaction phase
the states of the transport channels at each lattice site are randomly
permuted.\footnote{We should note that this approach can be considered
a field-theoretic molecular dynamics: the fields, not the particles,
are permuted. See Ref.~\cite{cambook} for an early discussion of this
distinction.}  The random permutation is applied to all lattice sites
simultaneously, with each permuted independently of all other sites.
Since the system is initialized with at most one particle in each
channel, this restriction is preserved by the dynamics. Thus all the
particles execute simultaneous random walks with exclusion. In the
model considered by Chopard and Droz, each channel at site $\x$ at
time $t$ was simultaneously permuted by a rotation of $r(\x,t)\cdot\pi
/2$ radians, where $r(\x,t)\in\{0,1,2,3\}$ is a random number with
values that occur with a frequency corresponding to a specified
probability distribution. By choosing this probability distribution
appropriately, we can control the mean-free path (i.e., the average
number of sites a particle advances before having its velocity
permuted).  Thus after the permutation,
\begin{equation}
N'_i(\x,t) = N_{i+r(\x,t)}(\x,t),
\end{equation}
where here and in what follows a sum in a subscript, such as
$(i+r(\x,t))$, is taken modulo four.  During the subsequent streaming
phase the particles are transported along the lattice:
\begin{equation}\label{eq:simple-stream}
N_i(\x,t+1) = N'_i(\x -\c_i, t) = N_{i+r(\x-\c_i,t)}(\x-\c_i,t).
\end{equation}

\begin{figure}[tbp]
\bigskip
\hfill{\hfill
\fbox{\resizebox{0.25\textwidth}{!}{
\includegraphics{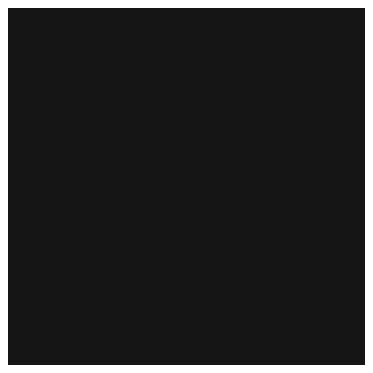}}}\hfill\hfill
\fbox{\resizebox{0.25\textwidth}{!}{
\includegraphics{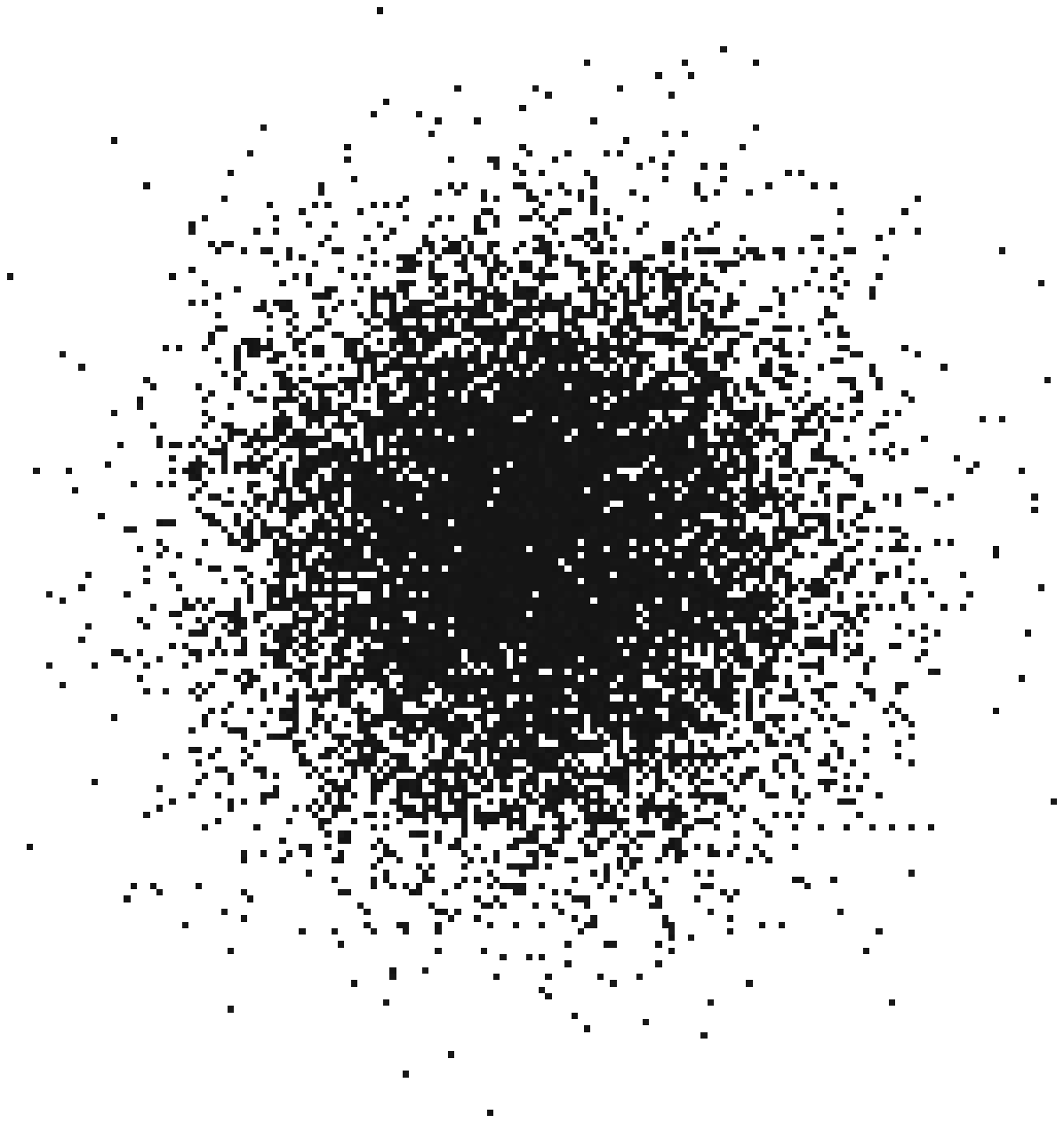}}}\hfill}\hfill
\bigskip
\caption{A realization of diffusion simulated by a simple CA rule
which conserves particle number (mass) but not momentum. Particles
move with unit velocity along the lattice directions of a
two-dimensional square lattice, of size $L\times L$ where $L=512$. For
each lattice site, at each update, the velocities of the particles
(1's) and holes (0's) at
that site are randomly permuted.  The random permutation used at each lattice site is
chosen independently.  Starting from an initial uniform block of particles, as
shown in (a), we observe the diffusive behavior shown in (b), which
corresponds to $t=360$ complete updates of the lattice.}
\label{fig:diffusion}
\bigskip
\end{figure}
Though Chopard and Droz considered a more general model, we restrict
ourselves to the case where $r(\x,t)$ is uniformly and independently
distributed amongst the four possible values for each $(\x,t)$.
Figure~\ref{fig:diffusion} shows the qualitative diffusive behavior of
the model.  The system is initialized with a uniform block of
particles occupying an otherwise vacant two-dimensional lattice, as
shown in Fig.~\ref{fig:diffusion}(a).  After $t=360$ updates of the
full lattice we observe the configuration shown in
Fig.~\ref{fig:diffusion}(b).

\subsubsection{Establishing the lattice {Boltzmann} equation}\label{subsubsec:LBA}
Up to here, we have described the dynamics for a single
realization of the lattice, as it is actually implemented and run on
a computer in practice.  However, for purposes of understanding how to
the use this technique as a realistic generator of diffusion, we would
like to know ``what happens on average in the limit'' (where these
terms have not yet been defined).  Figure~\ref{fig:diffusion} shows what
happens for a particular realization; in this section, we take the
average; and in the next section, we take the limit.

We first consider the mathematical idealization of an ensemble of all
such simulations.  We consider that each member of the ensemble is
initialized in a consistent way (as discussed in detail at the end of
this section), and evolves independently.  There are many
quantities one could average over this ensemble to test the model, but
the main quantity of interest is the state of each channel:
$N_i(\x,t)$.  Taking the expectation of Eq.~\ref{eq:simple-stream}
gives:
\begin{equation}\label{eq:ave}
\left<N_i(\x,t+1)\right> = \left<N_{i+r(\x-\c_i,t)}(\x-\c_i,t)\right>
			= \left<\frac{1}{4}\sum_{j=0}^3
				N_j(\x-\c_i,t)\right>
			= \frac{1}{4}\sum_{j=0}^3
				\left<N_j(\x-\c_i,t)\right>.
\end{equation}
Note that we used the fact that the expectation of a sum is the sum of
expectations, regardless of the correlations between distinct
channels.  Finally, defining the occupation number as
$n_i(\x,t)\equiv\left<N_i(\x,t)\right>$ gives the lattice-Boltzmann
equation\footnote{A lattice-Boltzmann equation (LBE) describes a
dynamical system spatially arrayed on a lattice but with continuous
variables at each site (as opposed to discrete variables, which is the
case for CA's). LBE's have many properties which make them better
suited for modeling hydrodynamic behavior.  Yet in situations where
microscopic noise, kinetic fluctuations and interparticle correlations
play an essential role, one must use LGA models.  For a discussion of
the striking difference in phenomenology accessible to LGA as opposed
to LBE see Ref.~\cite{bmb-LBE-amphifluids}.} for our system,
\begin{equation}\label{eq:cd-boltz}
n_i(\x,t+1) = \frac{1}{4} \sum_{j=0}^3 n_j(\x-\c_i,t).
\end{equation}
Averaging over the ensemble in this way transforms our discrete
Boolean variables into continuous probabilities.

Equation~(\ref{eq:cd-boltz}) is one of the few instances where an
interesting physical dynamics is exactly described by a linear
lattice-Boltzmann equation (as discussed above, correlations between
channels do not matter).  Typically, there are interactions between
channels, so in order to perform the ensemble average (as in
Eq.~(\ref{eq:ave})), one must make the approximation that the channels
are uncorrelated ({\em i.e.}, the assumption of molecular chaos), and
the resulting equations are nonlinear.  In the present case there are
no interactions between the channels because the streaming and
permutations are {\em data blind}, meaning they are not conditional on
the state of the channels.  As there is still confusion about the
correspondence between microscopic dynamics and the macroscopic limit,
especially in the context of the increase in macroscopic entropy which
may accompany a microscopically reversible
dynamics\cite{Lebowitz,stauffer-q2r}, models described by linear
equations may be illuminating. Exact solutions can be obtained and
discrepancies with the continuous limit discussed
rigorously\cite{ChopDroz-book}.

\subsubsection{Mapping onto the diffusion equation}\label{subsec:analytics}

If we now view each $n_i(\x,t)$ as a smooth function of a continuous
spacetime, we can derive a partial differential equation that governs
the dynamics of the total density, $\rho(\x,t)$, of particles at a
given site:
\begin{equation}\label{eq:rho}
\rho(\x,t) \equiv \sum_{i=0}^3 n_i(\x,t).
\end{equation}
Combining this with Eq.~(\ref{eq:cd-boltz}) we can write out the
evolution of the density,
\begin{eqnarray}\label{eq:dens-evolv}
\rho(\x,t+1) & = & \sum_{i=0}^3 n_i(\x,t+1) \nonumber \\
\ & = & \frac{1}{4}\sum_{i=0}^3 \left[ \sum_{j=0}^3 n_j(\x-\c_i,t)
\right] = \frac{1}{4} \sum_{i=0}^3 \rho(\x-\c_i,t).
\end{eqnarray}

The limit of interest is the diffusive regime where the time step,
$\Delta t$, and lattice spacing, $\Delta x = \left|c_i\right|$,
approach zero: $\Delta x \rightarrow 0$, and $\Delta t\rightarrow 0$,
while $\left(\Delta x\right)^2 /\Delta t\rightarrow$ constant.  Note
in this regime (of diffusion on an infinite lattice), the propagation
velocity, proportional to $\Delta x/\Delta t$, becomes infinite.  This
unphysical propagation velocity is actually a well known pathology of
the diffusion equation, Eq.~(\ref{eq:ficks}).

Taylor expanding the spatial terms in Eq.~(\ref{eq:dens-evolv}),
keeping those terms to order $\left(\Delta x\right)^2$, we obtain
\begin{eqnarray}\label{eq:taylor-space}
\rho(\x,t+1) & = & 
\frac{1}{4} \sum_{i=0}^3 \sum_{j=0}^3 \left[ 
n_j(\x,t) + \Delta x \left(\c_i \cdot \nabla\right) n_j(\x,t) +
\frac{(\Delta x)^2}{2} \left(\c_i \cdot \nabla\right)^2 n_j(\x,t)\right]
\nonumber \\
\ & = & \rho(\x,t) + \frac{\left(\Delta x\right)}{4}
\sum_{i=0}^3 \sum_{j=0}^3 (\c_i \cdot \nabla)n_j(\x,t)
+ \frac{\left(\Delta x\right)^2}{8} \sum_{i=0}^3 \sum_{j=0}^3 (\c_i
\cdot \nabla)^2 n_j(\x,t).
\end{eqnarray}
After noting that $\left(\c_i = - \c_{i + 2}\right)$ and
$\left[(\c_i \cdot \nabla)^2 + (\c_{i+1} \cdot \nabla)^2\right] =
\nabla^2$,  we can simplify Eq.~(\ref{eq:taylor-space}):
\begin{equation}\label{eq:taylor-space-simple}
\rho(\x,t+1) = \rho(\x,t) + \frac{\left(\Delta x\right)^2}{4} \nabla^2
\rho(\x,t). 
\end{equation}
Taylor expanding to order $\left(\Delta t\right)$ we obtain the 
diffusion equation:
\begin{equation}\label{eq:ca-ficks}
\partd{\rho(\x,t)}{t} = D \nabla^2 \rho(\x,t),
\end{equation}
where the diffusion constant,  $D= (\Delta x)^2 / 4\Delta t.$ 

We have shown above that the ensemble as a whole reproduces the
diffusion equation in the limit of an infinitely fine lattice and
continuous time.  However, for purposes of an actual LGA simulation,
all we have is a single member of the ensemble in a discrete
spacetime.  In particular, the state of each individual channel is a
$0$ or $1$ and not a real number between $0$ and $1$ as results from a
lattice-Boltzmann treatment.  So we want to argue that a typical
member of the ensemble---as generated by a single run of a
simulation---provides a realistic rendition of the diffusion of a
collection of particles.

The main thing we want in this respect is that the expected number of
particles in a patch of lattice sites is in some sense the same as the
expected number in the corresponding region of continuous space.  More
specifically, we demand equivalence at the mesoscopic scale (the scale
at which patches are large compared to the lattice spacing but small
compared to the characteristic scale of the system as a whole).  Note
that it is also at this mesoscopic scale that we demand equivalence
between LGA and lattice-Boltzmann equations.  We could sum
$\rho(\x,t)$, as obtained from the lattice-Boltzmann formulation, over
a patch of lattice sites and obtain a quantity representing the number
of particles in that patch.  On the other hand, we could generate
individual members of the ensemble from simulations according to the
LGA formulation, count the number of particles in the patch, and then
take the ensemble average of this count.  Both approaches give exactly
the same quantity.  Much as is the case for the separate channels
feeding into a single site, as discussed following Eq.~(\ref{eq:ave}),
the expectation over the ensemble for this count in the LGA
formulation is the same as the sum of the expectations giving
$n_i(\x,t)$ in the lattice-Boltzmann formulation.

This is true as long as each member of the ensemble is chosen to have
the appropriate average mesoscopic initial conditions.
Equation~(\ref{eq:cd-boltz}) provides an exact description of any
ensemble that starts with the correct average at each lattice site, no
matter how strong the correlations between sites. However, not every
member of such an initial ensemble is necessarily ``typical'' in that
it could be consistent with the mesoscopic conditions, yet have a very
unusual pattern of correlations between the cells; for example, the
particles might align to form an image of a face.  To prevent this in
practice, one initializes each {\em channel} at a relevant site
$(\x,t)$ independently at random to be a $1$ with a probability given
by $n_i(\x,t)$ and to be $0$ with a probability given by
$1-n_i(\x,t)$. This generates a sample of the maximum entropy ensemble
that is consistent with the initial specified averages.  Moreover, the
sample
is self-averaging in the sense that the average density of particles
in any patch around a point will closely match the initial average at
that point.  Finally, at any point in the simulation, the
configuration is likely to be a typical member of the ensemble because
the dynamics only serves to increase coarse-grained entropy, not
generate highly correlated patterns. It is these typical ensemble
members that provide the most realistic rendition of the diffusion of
a collection of particles.

While the above analysis is very promising, we should mention some
caveats of the model.  In particular, the simulation cannot match more
complicated statistics of a real physical system.  For example, the
fluctuations in the particle number are bounded above by the number of
channels in any patch whereas in principle, there is no upper bound in
a continuum.  Also, the specific microscopic details will never be the
same as in a real physical system or from one run to the next.  While
the exact configuration of all the particles do not matter to the
average density of the diffusing species alone, they do matter a great
deal when they are coupled to a system that is very sensitive to
initial conditions, such as the reversible aggregation model that will
be described in Sec.~\ref{subsec:ra}, where microscopic fluctuations
dictate the dynamics.

\subsection{The dimension-splitting technique}

\begin{figure}[tpb]
\bigskip
\hfill\hfill
\resizebox{0.65\textwidth}{!}{
\includegraphics{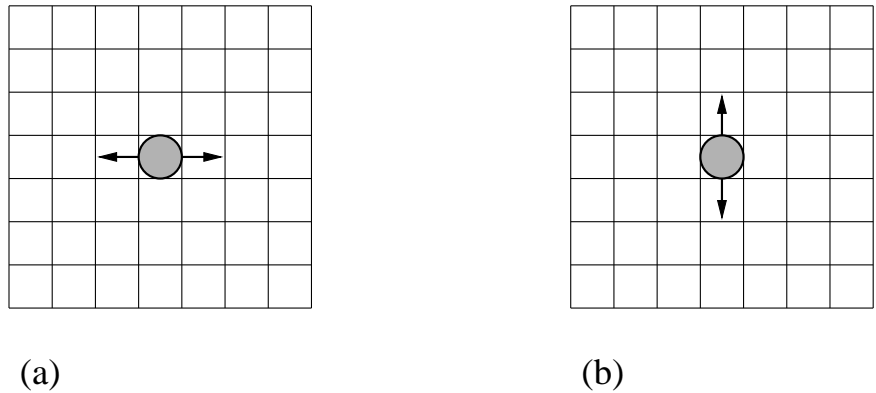}}
\hfill\hfill
\bigskip
\caption{An illustration of how the degrees of freedom are
reinterpreted using a partitioning scheme. A snapshot of a lattice
containing a single particle, represented by the shaded circle, is
shown.  The arrows represent the allowed directions of motion.  In (a)
the degrees of freedom are considered to be the directions of motion
along the horizontal axis. In (b) they are the directions along the
vertical axis.}
\label{fig:two-channels}
\bigskip
\end{figure}
Our approach is to consider the multi-dimensional diffusion process as
a sequence of one-dimensional diffusion events.  Each one-dimensional
diffusion step consists of an interaction ({\it i.e.}, mixing) phase
and a transport phase. We need consider only a single transport
channel and its opposite channel: $M_0(\x,t)$ and $M_1(\x,t)$.  First
consider the mixing phase.  An unbiased, random binary variable,
$\eta(\x,t)\in\{0,1\}$, $\left<\eta(\x,t)\right>=1/2$, is sampled to
determine whether the states of the two channels at site $(\x,t)$ are
interchanged:
\begin{eqnarray}
M'_0(\x,t) & = & [1-\eta(\x,t)] M_0(\x,t) + \eta(\x,t) M_1(\x,t) \nonumber
\\
M'_1(\x,t) & = & [1-\eta(\x,t)] M_1(\x,t) + \eta(\x,t) M_0(\x,t).
\end{eqnarray}
Then the updated channels are transported along the $\hat{x}$
direction, as shown in Fig.~\ref{fig:two-channels}(a):
\begin{eqnarray}\label{eq:x-trans}
M_0(\x,t) & = & M'_0(\x-\hat{x}, t-1) \nonumber \\
M_1(\x,t) & = & M'_1(\x+\hat{x}, t-1). 
\end{eqnarray}

Once these two substeps are complete, we consider diffusion along the
$\hat{y}$ direction which is also composed of the analogous mixing and
transport steps (i.e., substitute $\hat{y}$ for $\hat{x}$ in
Eq.~(\ref{eq:x-trans})).  As shown in Fig.~\ref{fig:two-channels}, we
first use the two channels to transport data along the horizontal
direction, then we use these same two channels to transport data along the
vertical direction. We thus reinterpret the function of the same two
degrees of freedom.

In this decomposition of a two-dimensional dynamics into two
one-dimensional dynamics, instead of four transport channels at each
site, we need only two. Thus at each lattice site, three bits of state
are required: two bits, $M_{\gamma}(\x,t)$ where $\gamma\in\{0,1\}$,
denote the presence or absence of a particle in each of the two
channels respectively; the third bit, $\eta(\x,t)$, is a binary
pseudorandom variable. Note, the conventional approach requires six
bits of state at each site.

The number of degrees of freedom interacting at any one time is
simplified (and remains so regardless of overall dimensionality).  Yet
we end up executing a sequence of four substeps (mix, then transport
along $\hat{x}$, mix then transport along $\hat{y}$). In contrast, the
full two-dimensional problem requires only two substeps (one mixing,
one transport).  Note also that extending the dynamics to one more
dimension requires only including an additional sequence of mixing
followed by transport ({\em i.e.,} an additional fractional timestep,
with the transport along the new dimension).\footnote{Note that when
dealing with diffusion on a two-dimensional triangular lattice we
would apply a sequence of three substeps, using the same two channels
to transport data subsequently along each of the three principle
lattice directions.  Generalization of
transport on a more complex higher-dimensional lattice would require
addition of a substep along each new principle lattice direction.}

\begin{figure}[tp]
\bigskip
\hfill
\resizebox{0.7\textwidth}{!}{
\includegraphics{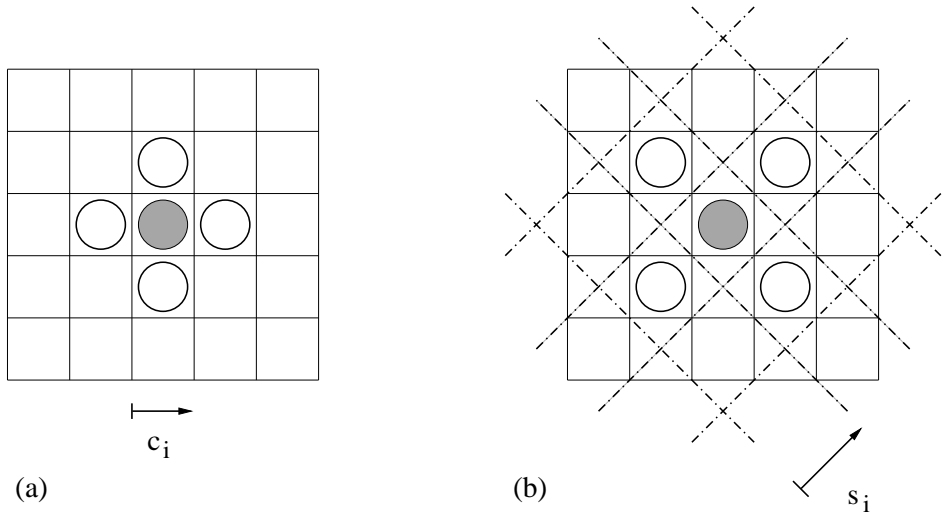}}
\hfill
\bigskip
\caption{A snapshot of the lattice. A single particle is shown as the
shaded circle. The potential sites that a particle can occupy at the
next iteration are shown as open circles.  (a) Updating using the
standard LGA diffusion algorithm.  Note the lattice spacing,
$\left|c_i\right|$.  (b) Updating using the dimension-splitting
technique. At the next iteration, the particle advances one step along
an oblique lattice.  This oblique lattice is shown superimposed on the
original lattice.  Note the lattice spacing, $\left|s_i\right| =
\sqrt{2}\left|c_i\right|$.}
\label{fig:dual-lattice}
\bigskip
\end{figure}

In the two-dimensional diffusion model considered, we find that at the
end of one complete update ({\em i.e.,} one update along $\hat{x}$ and
one update along $\hat{y}$) a particle has advanced one unit along
each dimension---one step along an oblique lattice with lattice spacing
$\left|\hat{s_i}\right| = \sqrt{2} \left|\hat{c_i}\right|$, where
$\left|\hat{c_i}\right|$ is the original lattice spacing.
The mapping onto this oblique lattice is shown in
Fig.~\ref{fig:dual-lattice}.

\subsubsection{Mapping onto the diffusion equation}

We show that this dimension-reduced approach to diffusion maps onto
the Chopard-Droz analysis discussed in
Sec.~\ref{subsubsec:define-model}, and thus also onto the diffusion
equation, Eq.~(\ref{eq:ficks}), in the appropriate regime. For
simplicity we again consider the two-dimensional case, and again deal
with the average over an ensemble of similar systems,
$m_i(\x,t)\equiv \left<M_i(\x,t)\right>$.

One complete update requires the four substeps discussed above. Thus
we consider the unit interval of time to be the time required to
complete these four substeps. The result at $t'= t + 1/2$ (of
transport and mixing along the horizontal direction) is
\begin{eqnarray}\label{eq:first-half}
m_0\left(\x,t+\frac{1}{2}\right) & = & \frac{1}{2}\left[
m_0(\x-\hat{x},t) + m_1(\x-\hat{x},t)\right] \nonumber \\
m_1\left(\x,t+\frac{1}{2}\right) & = & \frac{1}{2}\left[
m_0(\x+\hat{x},t) + m_1(\x+\hat{x},t)\right].
\end{eqnarray}
Now considering transport along the $\hat{y}$ direction we find that 
\begin{eqnarray}
m_0(\x,t+1) & = & \frac{1}{2}\left[ m_0(\x-\hat{y},t+\frac{1}{2}) +
m_1(\x-\hat{y},t+\frac{1}{2})\right] \\
\ & = & \frac{1}{4}\left[ m_0(\x-\hat{x}-\hat{y},t) +
m_1(\x-\hat{x}-\hat{y},t) + m_0(\x+\hat{x}-\hat{y},t) +
m_1(\x+\hat{x}-\hat{y}, t)\right]. \nonumber 
\end{eqnarray}
Analogously,
\begin{equation}
m_1(\x,t+1) = \frac{1}{4}\left[m_0(\x+\hat{x}+\hat{y},t) +
m_1(\x+\hat{x}+\hat{y},t) + m_0(\x-\hat{x}+\hat{y},t) +
m_1(\x-\hat{x}+\hat{y},t)\right].
\end{equation}
Let us redefine the lattice vectors such that $\s_0= \hat{x} +
\hat{y}$, $\s_1 = -\hat{x}+\hat{y}$, $\s_2 = -\s_0$, and $\s_3=-\s_1.$
Note that the unit vectors of the oblique lattice are longer than those
on the original lattice: $\left| \s_i \right| = \sqrt{2} \left| \c_i
\right|.$ The evolution of the density
\begin{eqnarray}
\rho(\x,t+1) & = & m_0(\x,t+1) + m_1(\x,t+1) \nonumber \\
\ & = & \frac{1}{4} \sum_{i=0}^3 \left[ \sum_{j=0}^1
m_j(\x-\s_i,t)\right]
= \frac{1}{4} \sum_{i=0}^3 \rho(\x-\s_i,t).
\end{eqnarray}
Comparing this equation to Eq.~(\ref{eq:dens-evolv}), we see that the
dimension-splitting model maps onto the conventional model since the
oblique lattice vectors, $\s_i$, map onto the original lattice
vectors, $\c_i$.  Reference~\cite{ChopDroz-diff} and
Ref.~\cite{ChopDroz-book} contain many quantitative plots of the
resulting diffusion.  Instead of reproducing those plots here, we
refer the reader to the original works. 

\subsubsection{Diffusion coefficients}\label{subsubsec:diff-coef}
We can easily control the coefficient of self-diffusion ($D$ in
Eq.~(\ref{eq:ca-ficks})), a property which will prove useful in the
models discussed in Sec.~\ref{sec:example-models}.  In particular, we
can increase the diffusion constant by a factor of $k^2$ in one of two
ways.  First we can transport the particles by $k$ lattice unit
vectors each streaming step.  Secondly, recalling that LGA models have
transport steps alternating with interaction steps, we can run $k^2$
diffusion steps for each particle interaction step.  With more effort,
we can use a non-uniform $r(\x,t)$ as shown by Chopard and
Droz\cite{ChopDroz-diff,ChopDroz-book} to increase or decrease the
probability of reversing direction at each interaction step and
thereby decrease or increase the mean free path, and hence the
diffusion constant.

\subsection{Constraints and conserved quantities}
As discussed above, the particles execute simultaneous random walks
with exclusion.  If we start with at most one particle per channel,
that constraint is preserved by the dynamics.  Moreover, the total
number of particles---alternatively viewed as mass, kinetic energy, or
total energy---is conserved.  Momentum is not conserved because
particles can change their states of motion without a compensating
change elsewhere.

Our model also has {\em spurious} conserved quantities (quantities
that we do not necessarily wish to conserve but are conserved
nonetheless).  The dynamics partitions the lattice into four
interleaved sublattices, and the number of particles on each
sublattice is independently conserved.  Consider a particle
initialized at the center of the lattice (marked by the shaded circle)
as shown in Fig.~\ref{fig:dual-lattice}(b).  At the next time step we
know it will occupy one of the sites marked with an open circle.
Likewise, only a particle initialized on a site marked by an open
circle will be able to occupy the site marked with the shaded circle
at the next time step.  If we think of superimposing a checkerboard on
the lattice, where we color the squares on one checkerboard sublattice
``red'' and those on the other ``black'', it is clear that all
particles on the red sites will occupy red sites at the end of any
complete update of the system.  Yet we can further divide each
checkerboard into two.  Again consider the situation illustrated in
Fig.~\ref{fig:dual-lattice}(b). A particle initialized on the shaded
site cannot occupy that shaded site at the next time step (even though
that shaded site is on the appropriate checkerboard). Thus only half of
the sites on that checkerboard are accessible in one time step, and so
we are simulating diffusion independently on each of four interleaved
sublattices.  In contrast, the Chopard-Droz model partitions the
lattice into two checkerboard sublattices, simulating diffusion
independently on each.\footnote{The original lattice-diffusion model
introduced by Toffoli uses a partitioning scheme which doesn't split
the space into uncoupled sublattices.}  If we increase the
dimensionality of our  model, we double the number
of independent checkerboards with each additional dimension. 

In Fig.~\ref{fig:diffusion}, we started all four sublattices in the
identical state, so the separate particle conservation on interleaved
sublattices is not apparent.  If we had started with one of the
sublattices occupied, at any point in the dynamics only one of the
sublattices would be occupied.  When diffusion is used as a component
of a larger simulation, these conservation laws are often broken by
interactions between subsystems. For example, as discussed below, we
consider interactions of diffusing vapor particles and a stationary
solid, where the interactions mix the sublattices.  A vapor particle
will aggregate regardless of the sublattice ordering, and subsequent
evaporation from the solid returns that vapor particle placed on an
arbitrary sublattice.  We can also eliminate the spurious conservation
laws in a more intrinsic way: we simply simulate a single sublattice
and embed it into the original lattice by shrinking it by a factor of
two in each direction. Odd and even timesteps must be embedded with
slightly different offsets (amounting to one-half the lattice
spacing), but this is just a slight complication in the
implementation.  For a detailed discussion of the embedding
technique see Ref.~\cite{nhm-spaceram}.


Often, when developing microscopic models, we wish to create models
which are microscopically invertible ({\em i.e.}, models which
conserve microscopic information).  If we use an invertible random
number generator for our source of random bits, we can build an
invertible model of diffusion, since we can then undo the mixing
permutation. (In the first example below, we use an independent invertible LGA as
the generator of a dynamical random number field).  We can always undo the
transport steps (by transporting in the opposite direction).  Thus
both the Toffoli/Chopard-Droz model and our model can be made invertible.

\section{Applications}\label{sec:example-models}

\subsection{The reversible aggregation model}\label{subsec:ra}
We developed this technique for simulating diffusion while building a
model of cluster growth via the aggregation of particles which were
previously diffusing along the lattice.  This model is an extension of
the athermal diffusion-limited aggregation (DLA)
model\cite{WittSand-dla}. We place a DLA lattice system in contact
with a simulated heat bath and allow only microscopically reversible
heat exchanges between the two subsystems.

During each aggregation event, a diffusing gas particle becomes a
stationary cluster member while latent heat is released. The heat is
explicitly modeled by introducing a diffusing heat particle onto a
heat bath lattice.  The inverse process is also allowed: a cluster
member can absorb a heat particle and evaporate, becoming a gas
particle.  Initially the heat bath is empty, and we observe rapid,
nonequilibrium growth.  During this initial phase the cluster
structures resemble those typical to DLA systems, such as the bushy
clusters formed as frost collects on a window pane.  During the
subsequent slow approach to thermodynamic equilibrium, the cluster
structures anneal until reaching the highest entropy macrostate
allowed for a connected cluster in a finite volume: a branched
polymer.  For more details of the reversible aggregation (RA) model
see Ref.~\cite{dsouza-nhm-RA}.

\begin{figure}[tbp]
\hfill
\fbox{\resizebox{0.35\textwidth}{!}{
\includegraphics{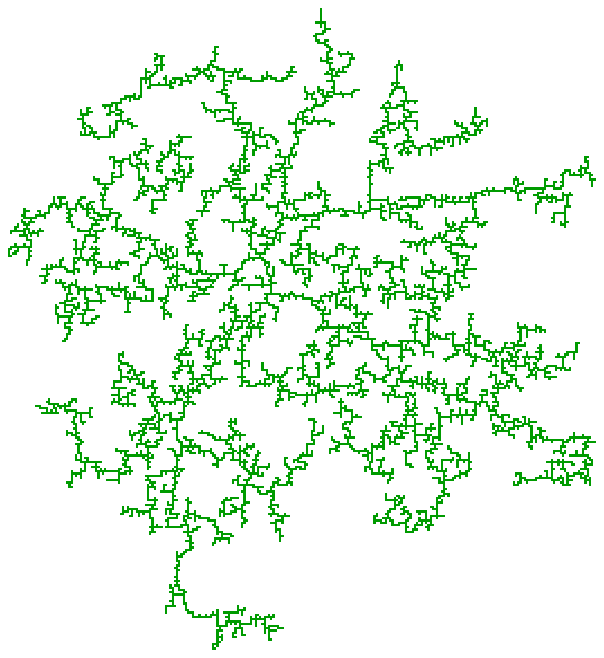}}}\hfill
\fbox{\resizebox{0.35\textwidth}{!}{
\includegraphics{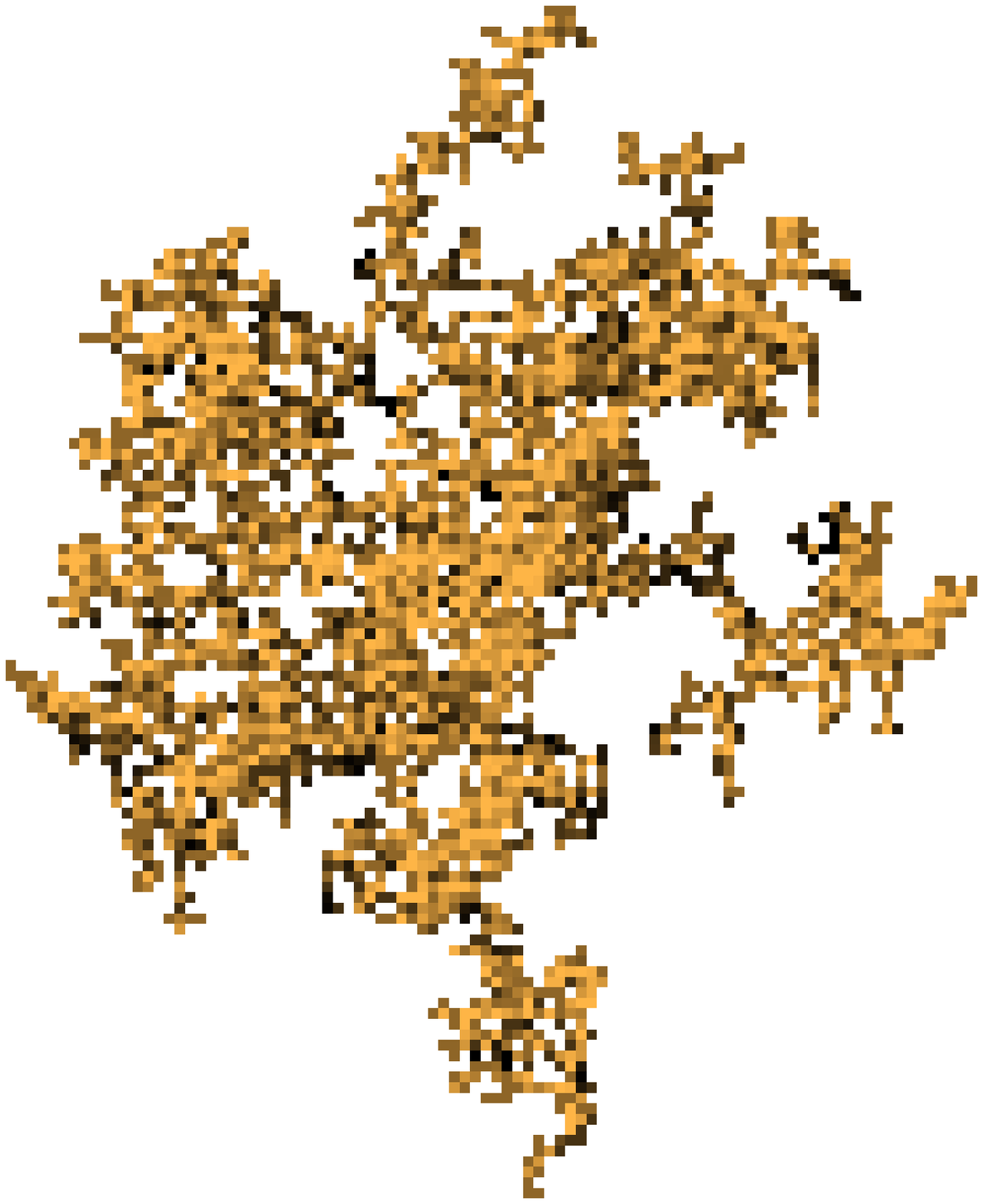}}}
\hfill
\bigskip
\caption{(a) A cluster grown via a two-dimensional RA
implementation. (b) A visualization of a three-dimensional RA cluster.
The rendering is done using an LGA algorithm which simulates discrete
light particles moving on the lattice\cite{cam8-waterloo}.  Diffusive
phenomena are at the foundation of the RA model: they are handled
via the dimension-splitting technique.  Note that extending the
two-dimensional model to a three-dimensional model required merely
adding one line of code.}
\label{fig:2d-3d-ra}
\bigskip
\end{figure}

This local, microscopic, reversible approach to modeling allows us to
explicitly track how information flow at the microscopic scale gives
rise to dissipation and properties such as increasing temperature at
the macroscopic scale.  We demonstrate that even far from
thermodynamic equilibrium, as the simulated clusters are growing and
annealing, we have a model with a well defined temperature, and that
these models can serve as a numerical laboratory for investigating
nonequilibrium thermodynamics.  Moreover, discrete reversible systems
such as this---which include local heat flow, creation of entropy, and
macroscopic dissipation---seem to capture the full essence of the
phenomena of pattern formation more completely than do irreversible
models.  For more details see Refs.~\cite{dsouza-nhm-RA,Dsouza-phd}.

Simulations of the RA model were implemented and run on the CAM-8
lattice-gas computer\cite{cam8-waterloo}.
Figure~\ref{fig:2d-3d-ra} shows a typical RA cluster, first for a
two-dimensional implementation, then for a three-dimensional one.
Implementing the three-dimensional model was a simple matter of
including one more line of code in our program (for transport along
the third dimension).  Note that the diffusion technique forms the
foundation of our model, which involves two interacting diffusion fields
(a field of gas particles and a field of heat particles).  We can
control the diffusion coefficient independently for each field, as
discussed in Sec.~\ref{subsubsec:diff-coef}, and observe the change
in the cluster structures and time to reach equilibrium as a function
of this parameter.

\subsection{A model for sedimentation}

Another application which makes use of the dimension-splitting
technique in a reaction-diffusion context is the Biofilm CA
model.  This is a model developed by Pizarro, Griffeath and
Noguera\cite{Pizarro-sludge,Pizarro-phd}, and programmed by Shalizi
and Margolus on the CAM-8\cite{cam8-waterloo}.

The Biofilm CA simulates the growth of biological films (e.g. sludge,
dental plaque) by having food and bacteria react and diffuse in two or
more dimensions.  The diffusion rate for the food depends on whether
there is bacteria present, and the model includes birth, food sources,
eating, death by starvation, random death, boundaries, attachment to a
substrate, attachment to an aggregate, death of unattached cells, and
the periodic removal of entire bacterial clusters which have become
disconnected from the substrate (this simulates the action of water
currents).

\begin{figure}[tbp]
\hfill
\fbox{\resizebox{0.35\textwidth}{!}{
\includegraphics{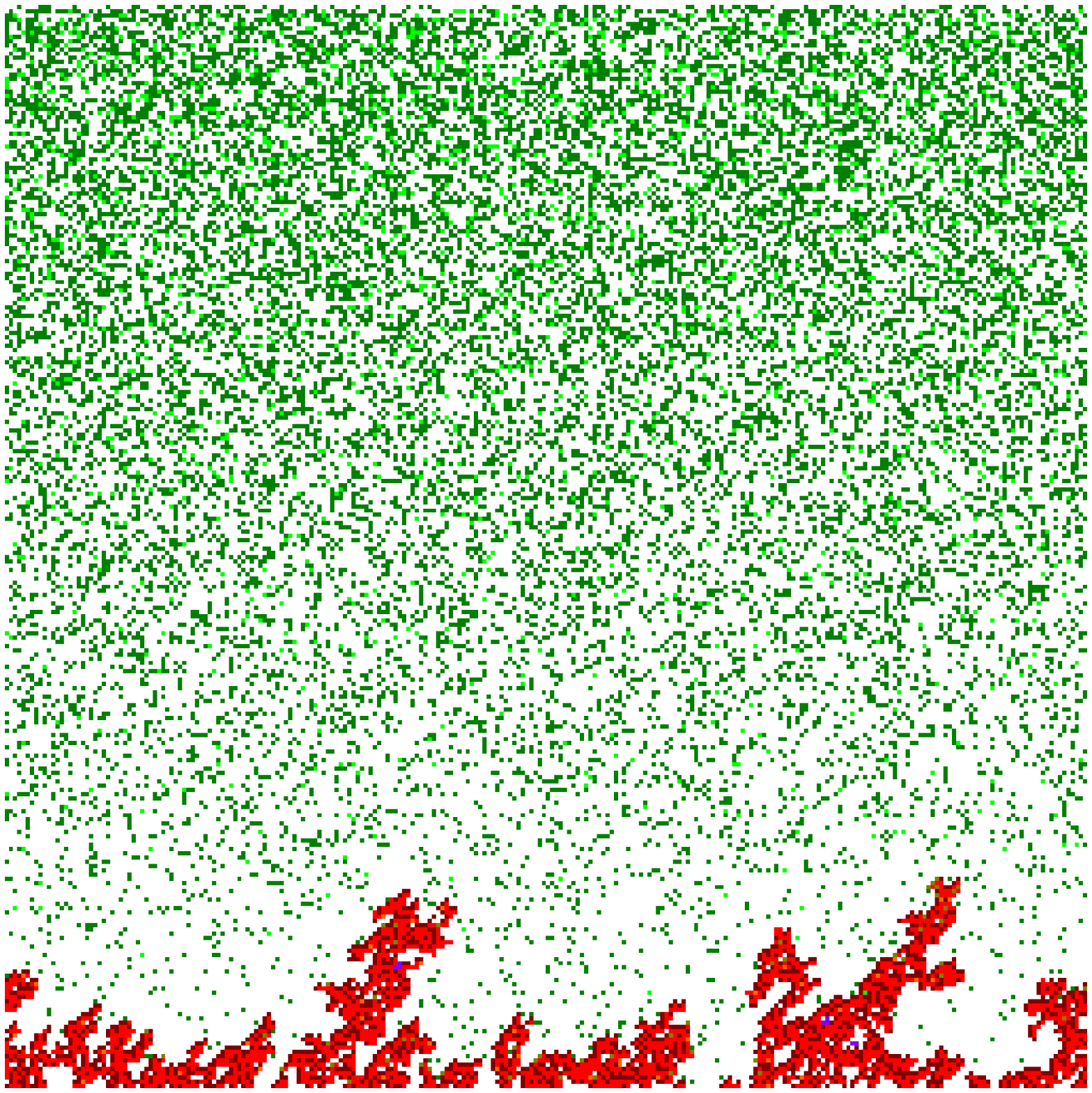}}}\hfill
\fbox{\resizebox{0.35\textwidth}{!}{
\includegraphics{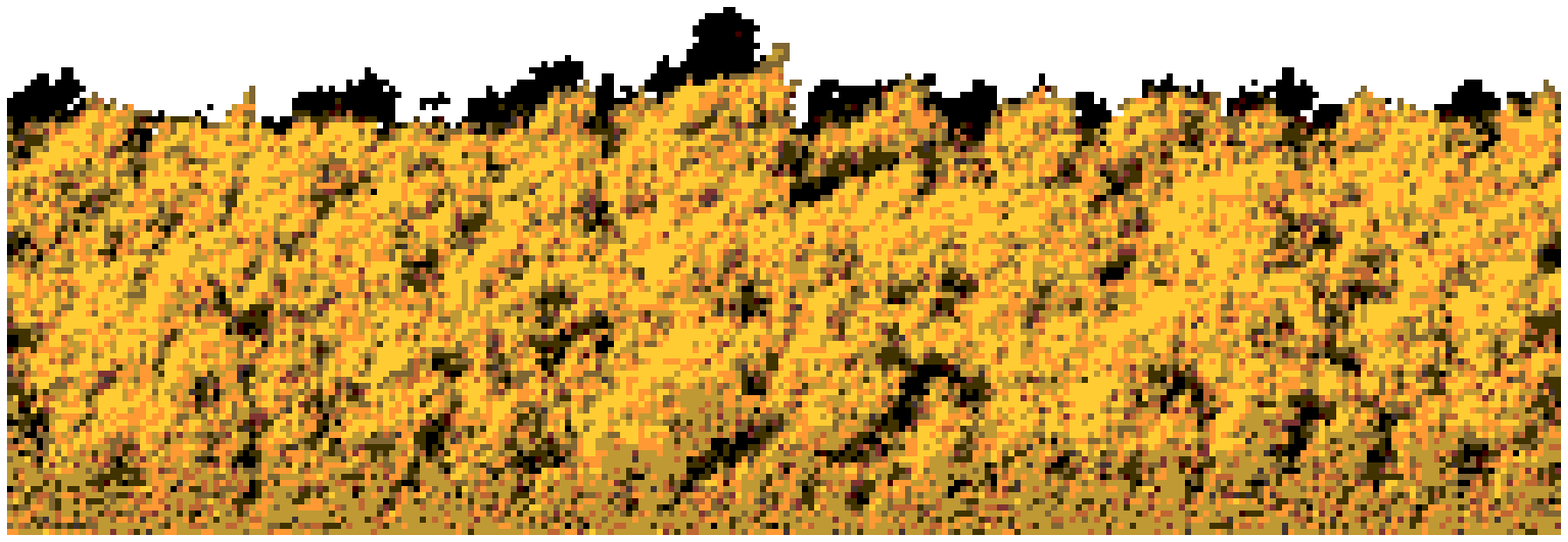}}}
\hfill
\bigskip
\caption{Dimension-splitting biofilm simulation.  (a) A
two-dimensional simulation run on a $256 \times 256$ space.  Food enters
at the top, and a biofilm grows on the substrate at the bottom.  (b) The
same simulation run in three dimensions on a $256 \times 256 \times 256$
lattice.  Rendering is done as in Fig.~\ref{fig:2d-3d-ra}(b). 
We're looking at a downward angle at the $256 \times 256$ substrate,
coated with biofilm growth.  The light source is on the left. The food
particles are not shown.}
\label{fig:biofilms}
\bigskip
\end{figure}

The diffusion in this model is handled one dimension at a time.  Each
diffusing species can occupy one or both (or neither) of two channels
at each lattice site, and the contents of the channels are alternately
swapped (or not swapped) at random, and then transported oppositely
along one dimension.  For the food particles, each channel contains a
single bit (present or not).  For the bacteria, each channel contains
two bits (present or not, and if present how hungry).  In regions
which should have a longer mean free path, the mixing probability is
reduced.  Food diffuses in from the top of the space, bacterial growth
starts on the substrate at the bottom.  To reduce the time taken for
food to initially reach the bacteria by diffusion, food starts off
uniformly distributed across the space.  For more details on the model
see Ref.~\cite{Pizarro-phd}.

The model was initially developed as a two-dimensional system, which
could be run quickly and easily, and conveniently visualized.  The
model was tested and debugged in two-dimensions until it behaved as
expected.  A typical state in the growth of a two-dimensional biofilm
is shown in Fig.~\ref{fig:biofilms}(a).  The model was then
generalized to three dimensions without adding any bits to each
lattice site or defining any new rules to apply to each site.  The
sequence of operations needed to update a single dimension was simply
iterated three times instead of two, with the third iteration
involving transport of diffusing particles along the added third
dimension.  An appropriate three-dimensional initial state was
prepared, and generic three-dimensional visualization routines were
activated.  The model immediately behaved as expected in
three dimensions---no further debugging was needed.  Only fine tuning
of initial densities, diffusion rates, etc., was required.  A typical
state in the growth of a three-dimensional biofilm is shown in
Fig.~\ref{fig:biofilms}(b).


\section{Discussion}

We have introduced a dimension-splitting technique for use in LGA
models with diffusive particle transport.  For an LGA of arbitrary
dimensionality, interactions involving only one dimension at a time
are sufficient for iterating the dynamics.  Thus, a model can be
developed in one dimension, and extended to more dimensions simply by
adding additional fractional timesteps.  The same dynamics are applied
in each fractional timestep, with the same diffusing particles moving
along each dimension in turn.  By reinterpreting and reusing the same
degrees of freedom and the same interactions at each fractional
timestep, diffusion is performed with a small amount of state
information.

Using this scheme, simple qualitative models involving
diffusion can easily be constructed using a kind of stylized molecular
dynamics.  We can appeal to our intuition in constructing such models,
simply including all relevant species and making them all interact in
a physically reasonable manner.  Constructing more realistic models
involves more work, however, since the diffusion behavior of our model
has only been demonstrated in the mesoscopic limit.  To use the model
only in this limit would involve running a very large number of
diffusion steps for each particle interaction step.  This might be
resolved by requiring a finer lattice for the diffusers than for other
entities in the simulation.  For many models,
however, we don't expect that a quantitatively realistic interaction
requires explicit mesoscopic averaging of the microscopic constituent
simulations.  Demonstrating realism for a given kind of simulation requires
both theoretical work and experimental verification.  Even models
which are not completely realistic, however, may be view as
``toy'' physical systems which can be simulated exactly and analyzed
theoretically\cite{dsouza-nhm-RA}. 

It seems that there is an important cost for our model's simplicity:
the dynamic range in particle density at each lattice site is small.
$N$ bits of state at each site yield a maximum density of only $N$
diffusers per site.  Numerical simulations of diffusion, in contrast,
allow up to $2^N$ diffusers to be represented at each site using the
same amount of state, with only a linear increase in the computational
complexity of the dynamics.  However, this contrast is only apparent.
The different species present at a lattice site can of course be given
different power-of-two weights: each ``1'' of species $A$ might represent a
single particle, each ``1'' of species $B$, two particles, etc.
Independent dimension-split diffusion of each such species provides a
large dynamic range while retaining exact particle conservation.

LGA models with as few as two bits of diffusing state at each lattice
site can be constructed using the dimension-splitting technique.  The
benefits of similar economies in simulation state have recently been
discussed in other contexts.  For example, in the context of
reversible compilers see Ref.~\cite{Perumala-rev-comp}; in the context
of systems relying on pseudorandom numbers see Ref.~\cite{Nisan92}.

It might seem that the operation on one dimension at a time would
result in slower simulations than a more complex LGA model, which
mixes data in several dimensions at once.  However, from a
computational complexity
standpoint, this is not at all obvious, since the more complex LGA
model would have more data to interact.  If, for example, diffusion is
being handled by circuitry capable of complex operations, then using
only simple operations may waste resources.  If, on the other hand,
only very simple operations are available in the circuitry, and only
one operation can be performed at a time, then more data would require
more operations and mean a slower simulation.  If it is ever possible
to perform simulations with molecular computers, the latter of
these two situations seems much more likely than the former.


Our dimension-splitting diffusion technique is particularly well
suited for use in concert with other dimension-splittable
primitives.  Composite models based only on such
primitives share the properties of reduced state
and the possibility of developing models in one or two dimensions, and
then running them without change in higher dimensions.

A useful primitive of this type is the LGA-fluid attraction and
repulsion technique used by Yepez and Appert in their
momentum-conserving lattice-gas aggregation
models\cite{yepez-blobs,apert-blobs}.  In these models, pairs of
particles lying along principle lattice directions and separated by
a given distance are brought together to interact.  By using a set of
distances and specified types of interactions (repulsive interactions
turn both particles away from their midpoint, attractive interactions
turn them towards it), prescribed potentials can be synthesized
operating between particle aggregates.\footnote{Their
original models didn't use dimension-splitting but instead used
``virtual particles'' moving in all directions at once to carry the forces.
In translating their models to the CAM-8 machine, we employed
dimension-splitting.}

Since these models aren't hydrodynamic, momentum conserving gas
interactions used to mix particles going in different directions can
also be made to interact one dimension at a time, forming a rest
particle from each head-on collision which then decays into two new
particles moving out in opposite directions---when the particle decays
can be determined stochastically.  Using the technique of the present
paper, diffusing components can also be included in such models.  Note
that in all of these models, the fractional timesteps are applied
along each of the principle lattice directions, so that in a
two-dimensional triangular lattice model, for example, there are three
transport steps rather than just two.  Thus a better name for the
technique might be ``lattice-splitting'' rather than
``dimension-splitting''.  It would be interesting to find additional
examples of such splitting techniques which can be used as components
of LGA models.


\bibliographystyle{unsrt}
\bibliography{/home/raissa/Bibl/ca,%
/home/raissa/Bibl/heat,%
/home/raissa/Bibl/dyn-ising,%
/home/raissa/Bibl/ca_rev,%
/home/raissa/Bibl/information,%
/home/raissa/Bibl/statmech,%
/home/raissa/Bibl/dla,
/home/raissa/Bibl/polymer,%
/home/raissa/Bibl/reac_diff,%
/home/raissa/Bibl/nanotech,%
/home/raissa/Bibl/xtalgrowth,%
/home/raissa/Bibl/rng,%
/home/raissa/Bibl/rev-comp,%
/home/raissa/Bibl/lgas,%
/home/raissa/Bibl/rwalks}

\end{document}